\documentclass[reprint, superscriptaddress, amsmath, amssymb, aps, prl]{revtex4-1}

\usepackage{xcolor}
\usepackage{graphicx}
\usepackage{dcolumn}
\usepackage{bm}
\usepackage{braket}
\usepackage{hyperref}
\hypersetup{colorlinks,
        linkcolor={blue!75!black!80!yellow},
        citecolor={blue!75!black!80!yellow},
        urlcolor={blue!75!black!80!yellow}
        }
\usepackage[compat=1.1.0]{tikz-feynman}
\usepackage[export]{adjustbox}

\usepackage{newtxtext}
\usepackage[cmintegrals]{newtxmath}
\usepackage{physics}

\usepackage[caption=false]{subfig}
\usepackage{adjustbox}

\newcommand{\Stanford}{Department of Materials Science and Engineering, Stanford University, Stanford, CA, 94305, USA}
\newcommand{\Yale}{Department of Materials Science, Yale University, New Haven, CT, 06520, USA}
\newcommand{\SIMES}{Stanford Institute for Materials and Energy Sciences, SLAC National Accelerator Laboratory, Menlo Park, CA, 94025, USA}

\begin{document}

\preprint{APS/123-QED}

\title{Exciton-polaritons and exciton localization from a first-principles interacting Green's function formalism}

\author{Zachary N. Mauri}
\thanks{These authors contributed equally}
\author{Christopher J. Ciccarino}
\thanks{These authors contributed equally}
\author{Jonah B. Haber}
\affiliation{\Stanford}
\author{Diana Y. Qiu}
\affiliation{\Yale}
\author{Felipe H. da Jornada}
\email{jornada@stanford.edu}
\affiliation{\Stanford}
\affiliation{\SIMES}

\date{\today}

\begin{abstract}
Exciton-polaritons -- hybrid states of photons and excitons --  offer unique avenues for controlling electronic, optical, and chemical properties of materials.
However, their modeling is mostly limited to formalisms that wash out atomistic details and many-body physics critical to describing real systems.
Here, we present an \textit{ab initio} Green’s function formalism based on the Bethe-Salpeter equation (BSE) wherein exciton-polaritons naturally emerge through an attractive, dynamical electron-hole exchange interaction.
In MgO and crystalline pentacene, this attractive interaction dramatically reduces exciton Bohr radii and increases, by one order of magnitude, transition dipole moments of exciton-polaritons.
Our calculations are in good agreement with experimental polariton dispersions in wurtzite CdS, and allow one to capture how electronic and collective excitations in materials are qualitatively modified through polaritonic effects.
\end{abstract}

\maketitle

\section{Introduction}\label{sec1}

Excitons are charge-neutral quasiparticles composed of bound electron-hole pairs whose interaction with electromagnetic radiation has garnered significant interest as a promising new pathway for manipulating the optoelectronic properties of materials. When the center-of-mass (COM) crystal momentum $\mathbf{Q}$ of an exciton is small enough such that its energy is nearly equivalent to that of a photon with the same $\mathbf{Q}$, the distinction between light and matter blurs, as strong coupling induces the formation of hybrid states known as (exciton-) polaritons~\cite{P58,H58}. In bulk materials, this strong coupling is responsible for a renormalization of the exciton dispersion and a surprising reduction of the optical absorption in clean samples~\cite{GKK11, K01, FMW71}. The mixed excitonic and photonic character of these quasiparticles makes them promising candidates for the realization of high-temperature Bose-Einstein condensates, Cooper pair binding agents, exotic topological phases, and a host of other nontrivial phenomena~\cite{KRK06, LKS10, KBC15, KHE18}.

Currently, there are several successful \textit{ab initio} approaches and computational methods to study excitons and their associated dispersion relations and electron-hole amplitudes. In particular, interacting Green's function formalisms based on the GW plus Bethe-Salpeter equation (BSE) approach can access the excitonic properties of materials ranging from bulk to nanostructured systems~\cite{HL86,RL00}. Recent years have also seen the extension of such formalisms to include the scattering of excitons due to bosonic modes such as phonons, as well as the formation of complex multiparticle excitations due to the coupling of excitons with a host of other quasiparticles~\cite{RQLN18, RJLN17, AL22, CR19}. Still, these first-principles frameworks have largely neglected the coupling with transverse photons which is responsible for the emergence of exciton-polaritons. Instead, most approaches that seek to capture polaritonic physics fit first-principles GW-BSE results to simplified electromagnetic models~\cite{BRQ86,KO10, LLP19} or include excitonic effects by dressing photons with an effective dielectric tensor. A notable exception is the work by S. Latini \textit{et al.}~\cite{LRGHR19}, who, focusing on excitons with zero center-of-mass (COM) momentum, diagonalized the coupled exciton-photon Hamiltonian in a basis of BSE exciton-photon product states to arrive at polariton energies and expansion coefficients, in an approach spiritually similar to Hopfield's theory of polaritons~\cite{H58}. Despite the success of these approaches in capturing the dispersion relation of polaritons, they are largely restricted to describing the photonic component of polaritons and cannot easily scale to capture polaritonic effects in structurally complex materials that display nontrivial relationships between exciton COM momentum and many-body exchange physics, including low-dimensional materials and molecular crystals~\cite{QCL15,QCNR21}.

In this work, we present a first-principles formalism to capture exciton-polaritons within a computationally efficient GW-BSE framework. We achieve this by replacing the instantaneous Coulomb potential in the electron-hole exchange interaction of the BSE with a dynamical, fully-relativistic bare photon propagator, leading to \emph{uncoupled} Dyson equations for the dressed exciton and photon propagators. Polariton dispersions are obtained from the poles of these propagators, while their residues encode the relative light-matter contributions of these composite states. We show that the matter component of polaritons is dramatically altered as these states become light-like, and is accompanied by a surprisingly large renormalization of the exciton Bohr radius~\cite{K01}. We understand this as the result of an effective electron-hole \emph{attraction} due to a dynamical exchange interaction, rather than the typical electron-hole \emph{repulsion} from a static exchange interaction. Our calculations reveal that this effect can qualitatively change the nature of excitons in prototypical bulk semiconductors such as crystalline pentacene, wherein retardation effects enable a transition from a Wannier- to a Frenkel-like polariton wavefunction, which cannot be captured with simpler continuum models. We anticipate such knowledge to be critical to polariton lasing, condensate formation, and thermalization in cavity systems.

\section{Theoretical formalism}\label{sec2}

The BSE allows one to solve for the interacting electron-hole correlation function~\cite{S88,RL00}, typically cast into an effective eigenvalue problem $H_\mathrm{BSE} \ket{S_\mathbf{Q}} = \Omega_\mathbf{Q}^S \ket{S_\mathbf{Q}}$, where $\ket{S_\mathbf{Q}}$ is an exciton with principal quantum number $S$ and COM wavevector $\vb{Q}$, $\Omega_\mathbf{Q}^S$ is the exciton excitation energy, and $H_\mathrm{BSE}$ is the BSE effective Hamiltonian. This equation is often solved in a basis of quasi-electrons and quasi-holes, in the Tamm-Dancoff approximation,
\begin{multline}
\label{eq:EigL}
    \left( E_{c\mathbf{k+Q}} - E_{v\mathbf{k}} \right) A_{vc\mathbf{kQ}}^S + \\
    \sum_{v'c'\mathbf{k}'} \Braket{vc\mathbf{k}\mathbf{Q} | K | v'c'\mathbf{k}'\mathbf{Q}} A_{v'c'\mathbf{k}'\mathbf{Q}}^S = \Omega_{\mathbf{Q}}^S A_{vc\mathbf{k}\mathbf{Q}}^S,
\end{multline}
where $A_{vc\mathbf{k}\mathbf{Q}}^S$ are expansion coefficients for describing excitons in terms of noninteracting electron-hole pairs with valence band $v$, conduction band $c$ and wavevector $\vb{k}$, such that $\ket{S_\mathbf{Q}}=\sum_{vc\mathbf{k}} A_{vc\vb{kQ}}^S \ket{v\vb{k},c\vb{k+Q}}$, and $E_{v\mathbf{k}}$ and $E_{c\mathbf{k+Q}}$ are the quasiparticle energies for a valence and conduction state, respectively.

\begin{figure}
    \centering
    \includegraphics[width=\columnwidth]{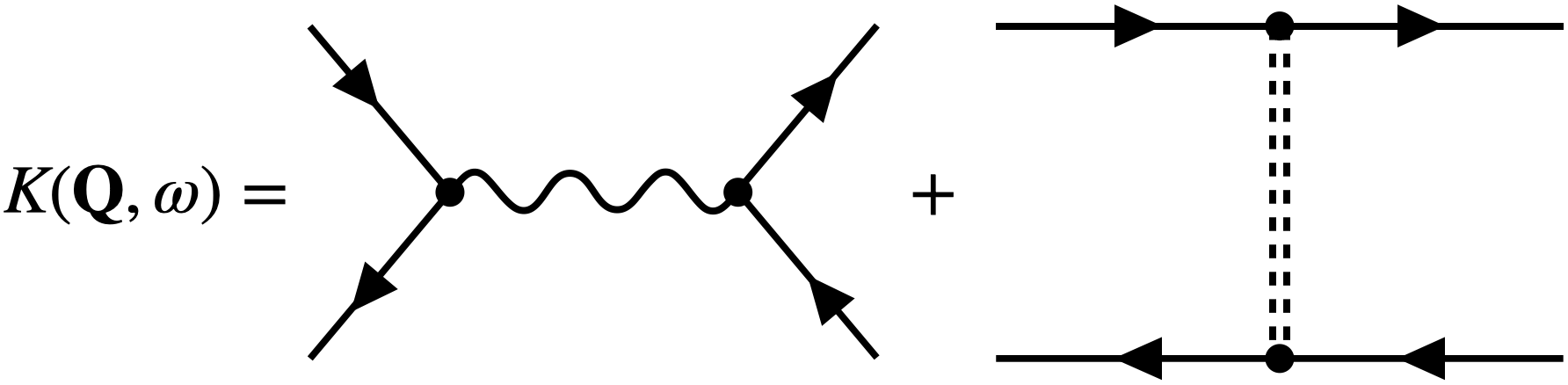}
    \caption{Diagrammatic representation of the Bethe-Salpeter equation (BSE) electron-hole interaction kernel $K$. The first and second diagrams on the right-hand-side are the electron-hole exchange interaction $K^\mathrm{x}$, and the direct interaction $K^\mathrm{d}$, respectively. Upper (lower) straight lines correspond to an electron (hole), the wavy line is a bare photon, and the dashed line is a screened Coulomb potential.}
    \label{fig:kernel_diagram}
\end{figure}

Considering only first-order interactions, the electron-hole interaction kernel $K(\mathbf{Q},\omega) = K^\mathrm{x}(\mathbf{Q},\omega) + K^\mathrm{d}(\mathbf{Q},\omega)$ is a sum of exchange $(K^\mathrm{x})$ and direct $(K^\mathrm{d})$ contributions, expressed diagrammatically in Fig.~\ref{fig:kernel_diagram}.
For most practical uses, both interactions are approximated by their \emph{static} $\omega\to0$ limits, resulting in exchange and direct interaction matrix elements
\begin{multline}
\label{eq:Kx}
    \Braket{vc\mathbf{kQ} | K^\mathrm{x} | v'c'\mathbf{k'Q}} = \\
    \frac{1}{V}\sum_{\mathbf{G}\ne\vb{0}} \rho_{cv}(\mathbf{k,Q,G}) v(\mathbf{Q+G}) \rho_{c'v'}^*(\mathbf{k',Q,G})
\end{multline}
and
\begin{multline}
\label{eq:Kd}
    \Braket{vc\mathbf{kQ} | K^\mathrm{d} | v'c'\mathbf{k'Q}} = \\ 
    -\frac{1}{V} \sum_\mathbf{GG'} \rho_{cc'}^*(\mathbf{k+Q,q,G}) W_\mathbf{GG'}(\mathbf{q}) \rho_{vv'}(\mathbf{k',q,G'}),
\end{multline}
where $\mathbf{G}$ is a reciprocal lattice vector, $\rho_{nn'}(\mathbf{k,q,G})=\sum_\alpha \Braket{n\mathbf{k+q},\alpha|e^{i(\mathbf{q+G)\cdot r}}|n'\mathbf{k},\alpha}$ are plane-wave charge density matrix elements with the spinor component $\alpha$ of the single-particle spinor wavefunctions summed over, $V$ is the crystal volume, $\mathbf{q} = \mathbf{k}-\mathbf{k}'$, and $W$ and $v$ are the statically screened and unscreened Coulomb interaction, respectively. The long-range ($\vb{G}=\vb{0}$) contribution to the exchange interaction is typically omitted when computing the macroscopic dielectric function of materials~\cite{DF84}.

Previous efforts to go beyond this static picture using a frequency-dependent kernel have focused on the direct interaction, where the relevant energy scale is determined by the difference between the exciton excitation energy and QP gap, $\Omega^S_\mathbf{Q} - (E_{c\mathbf{k+Q}} - E_{v\mathbf{k}})$. For most semiconductors and insulators, this difference is much smaller than the plasma energy $\omega_p$, which roughly dictates the onset of strong dynamical effects in the screening. Hence, for many materials a static approximation to the direct interaction is adequate~\cite{RL00,MS03}.

Unlike the direct interaction, the virtual photons mediating electron-hole exchange carry an energy $\Omega^S_\mathbf{Q}$ equal to that of the collective excitation one is solving for, \textit{i.e.}, Eq.~(\ref{eq:EigL}). At small values of $\mathbf{Q}$, this energy scale coincides with the dispersion $\omega = c |\mathbf{Q}|$ of physical transverse photons, whose coupling to excitons has been wholly omitted within the BSE as a consequence of the static approximation. We posit that the inability of the BSE to capture exciton-polaritons is in no way fundamental, and arises solely from this choice to treat the exchange interaction statically, effectively restricting the theory to unphysical time-like and longitudinal photons. To correct for this and properly account for the role that transverse photons have in mediating electron-hole exchange, we propose a fully-relativistic exchange interaction which treats excitons and photons on an equal footing.

\subsection{Relativistic exciton exchange interaction}

We expect retardation effects in $K^\mathrm{x}$ to capture polaritonic physics within the BSE; however, such dynamical effects would naively render BSE calculations impractical as one needs to diagonalize a large effective Hamiltonian $H_\mathrm{BSE}$ for each frequency and wavevector. Fortunately, we can simplify this process by first splitting the Fourier series expansion of Eq.~(\ref{eq:Kx}) into long-range ($\mathbf{G}=0$) and short-range ($\mathbf{G}\neq0$) components. In practical calculations of materials, any non-zero reciprocal lattice vectors will satisfy $\mathbf{G}\gg\omega/c$ for small $\mathbf{Q}$, so the short-range components can be safely approximated as instantaneous, allowing us to restrict dynamical effects solely to the long-range (LR) exchange interaction. 

We now work in the instantaneous BSE eigenbasis, or exciton basis $\{\ket{S_\mathbf{Q}}\}$, from solving the BSE in Eq.~\ref{eq:EigL} without dynamical effects and any LR exchange interactions.
In such a basis, the \emph{retarded} LR exchange interaction is
\begin{multline}
    \label{eq:RetardedKx}
    \Braket{S_\mathbf{Q} | K^\mathrm{x}_\mathrm{LR} (\omega) | S'_\mathbf{Q}} = \\
    \frac{1}{Vc^2}j^\mu_S(\mathbf{Q},\omega) D^{(0)}_{\mu\nu}(\mathbf{Q},\omega) j^{\nu\ast}_{S'}(\mathbf{Q},\omega),
\end{multline}
where $j^\mu_S$ is a current four-vector, $D^{(0)}_{\mu\nu}$ is the bare photon propagator, whose form is dependent on the choice of electromagnetic gauge.
Latin indices label spatial components (1,2,3); Greek indices label both time (0) and spatial components. Using a $\mathbf{Q}\cdot\mathbf{v}$ expansion of plane-wave matrix elements, with the velocity operator defined as $\hat{\mathbf{v}} = i[H_\mathrm{BSE},\hat{\mathbf{r}}]$ (see SI~\cite{SI}), the current four-vector matrix elements evaluates to
\begin{equation}
\label{eq:j0}
    j^\mu_S(\mathbf{Q},\omega) = 
    \begin{cases}
        e\frac{c\mathbf{Q}\cdot\mathbf{v}_S}{\Omega^S_\mathbf{Q}} & \mu = 0\\
        e\frac{\omega \hat{\mathbf{e}}^i\cdot\mathbf{v}_S}{\Omega^S_\mathbf{Q}} & \mu = i = 1,2,3
    \end{cases}
\end{equation}
where $\Omega^S_\mathbf{Q}$ is the energy of exciton $S$ obtained from the instantaneous Hamiltonian $H_\mathrm{BSE}$, $\mathbf{v}_S = \Braket{0|\hat{\mathbf{v}}|S}$ is the corresponding velocity matrix element, $e$ is the elementary charge, and $\hat{\mathbf{e}}^i$ are orthogonal unit vectors. We note that $j^\mu_S$ is defined for arbitrary on- and off-shell frequencies, and its precise form plays the somewhat subtle but crucial role of ensuring gauge invariance by satisfying the Ward identity~\cite{peskin_schroeder} $(\omega/c)j^0=\mathbf{Q}\cdot \mathbf{j}$ (see SI~\cite{SI}).

We note the two qualitatively different forms that Eq.~(\ref{eq:RetardedKx}) may take due to the $\mathbf{Q}\cdot\mathbf{v}$ dependence on the current four-vectors. For longitudinal ($\mathbf{Q} \parallel \mathbf{v}_S$) excitons, the time-like and spatial elements of $j^\mu_S$ are both non-zero but conspire to destroy the relativistic effects of Eq.~(\ref{eq:RetardedKx}) such that the diagonal exchange matrix elements for $S_\mathbf{Q}=S'_\mathbf{Q}$ yield a finite and constant shift equal to $\frac{4\pi e^2}{V} |\mathbf{v}_{S}/\Omega^S_\mathbf{Q}|^2$ (see SI~\cite{SI}). This result exactly matches that of the instantaneous exchange interaction in Eq.~(\ref{eq:Kx}) when a similar $\mathbf{Q}\cdot\mathbf{v}$ expansion of plane-wave matrix elements is applied~\cite{QCNR21,ABQ88}, and is a direct consequence of the unphysical nature of time-like and longitudinal photons which together mediate the instantaneous Coulomb interaction. 

In contrast, the diagonal exchange matrix elements of transverse ($\mathbf{Q} \perp \mathbf{v}_S$) excitons are equal to $\frac{4\pi e^2 \omega^2}{V(\omega^2 -|c\mathbf{Q}|^2)} |\mathbf{v}_S/\Omega^S_\mathbf{Q}|^2 $.
More generally, any photon with a polarization vector $\epsilon^\mu$ that has a nonzero component orthogonal to its four-momentum $Q^\mu=(\omega/c,\mathbf{Q})$ will undergo a dynamical coupling to excitons.
This is in agreement with the familiar result that only transverse excitons, or analogously transverse ($Q^\mu \perp \epsilon^\mu$) photons hybridize to form polaritons \cite{Davydov,ABQ88}. The exchange interaction's dependence on the orientation of $\mathbf{Q}$ relative to $\mathbf{v}_S$ has been well-described for cubic crystals in the instantaneous formulation of the BSE, where it is known as longitudinal-transverse (LT) splitting, and in studies of excitons in molecular crystals as Davydov splitting~\cite{D48}. As we will later show in the case of wurtzite CdS, this phenomenon is also responsible for the birefringence of optically anisotropic crystals.

\subsection{Retarded Bethe-Salpeter equation}

Having derived the retarded LR exchange interaction matrix elements, we may now introduce the main result of this work: the retarded Bethe-Salpeter equation (rBSE). Written in the exciton basis as a generalized eigenvalue problem, the rBSE is given by
\begin{multline}
\label{eq:RetardedEigL}
    \Omega^S_\mathbf{Q} A^P_{S\mathbf{Q}} + \sum_{S'} \Braket{S_\mathbf{Q} | K^\mathrm{x}_\mathrm{LR}(\Omega^P_\mathbf{Q}) | S'_\mathbf{Q}} A^P_{S'\mathbf{Q}} = \Omega^P_\mathbf{Q} A^P_{S\mathbf{Q}},
\end{multline}
and can be thought of as the polaritonic analog of Eq.~(\ref{eq:EigL}) for energies $\Omega^P_\mathbf{Q}$ and amplitudes $A^P_{S\mathbf{Q}}$ that correspond to polariton eigenstates with single-exciton components $\ket{P_\mathbf{Q}}=\sum_S A^P_{S\mathbf{Q}}\ket{S_\mathbf{Q}}$. Since the rBSE effective Hamiltonian is frequency-dependent and must be solved self-consistently for each polaritonic state independently, it is advantageous to instead work directly with the interacting, retarded electron-hole correlation function
\begin{equation}
\label{eq:Linv}
    L(\mathbf{Q},\omega) = \left[(\omega - \Omega^S_\mathbf{Q} + i0^+)\delta_{SS'} -  K^\mathrm{x}_\mathrm{LR}(\mathbf{Q},\omega)\right]^{-1},
\end{equation}
whereby $K^\mathrm{x}_\mathrm{LR}(\mathbf{Q},\omega)$ now appears as the proper exciton-polariton self-energy whose matrix elements are given by Eq.~(\ref{eq:RetardedKx}). In the vicinity of the light cone where polaritonic effects are most pronounced, the bare excitons used to construct the rBSE are nearly dispersionless. Making the approximation that $\Omega^S_\mathbf{Q} \approx \Omega^S_\mathbf{0}$, we can simplify our procedure by neglecting the $\mathbf{Q}$ dependence of the instantaneous BSE and diagonalize Eq.~(\ref{eq:EigL}) only at $\vb{Q}=\vb{0}$.

The correlation function $L(\mathbf{Q},\omega)$ is the most fundamental quantity of the rBSE and satisfies the well-known properties of Green's functions~\cite{Economou}, namely that it has poles at polariton eigenvalues $\Omega^P_\mathbf{Q}$ with residues that correspond to the fraction of the polariton composed of a single exciton. Polariton dispersions $\Omega^P_\mathbf{Q}$ are obtained by evaluating the exciton spectral function $A_\mathrm{xct}(\mathbf{Q},\omega) = -\frac{1}{\pi}\mathrm{ImTr}[L(\mathbf{Q},\omega)]$ on a dense grid of $\mathbf{Q}$ and $\omega$ values. For each point on this grid, the correlation function may naively be computed according to Eq.~(\ref{eq:EigL}) via the inversion of a matrix with dimensions equal to the number of all possible $v\rightarrow c$ transitions at each $\mathbf{k}$. Fortunately, the low-rank structure of the retarded exchange interaction allows us to express this inversion, using Woodbury's formula (see SI~\cite{SI}), as
\begin{multline}
\label{eq:Woodbury}
    L(\mathbf{Q},\omega) = \tilde{L}(\omega) + \\
    \frac{1}{Vc^2} \tilde{L}(\mathbf{Q}, \omega) j^\mu_S(\mathbf{Q},\omega)D_{\mu\nu}(\mathbf{Q},\omega)j^{\nu\ast}_{S'}(\mathbf{Q},\omega) \tilde{L}(\omega).
\end{multline}
It should be stressed that this inversion procedure is exact and can be physically interpreted as a dressing of the instantaneous electron-hole correlation function $\tilde{L}(\omega) = \delta_{SS'}/(\omega - \Omega^S_\mathbf{Q} + i0^+)$ by the full (\textit{i.e.}, improper) polariton self-energy. This is in essence equivalent to replacing the bare photon propagator $D^{(0)}$ that mediates the retarded exchange interaction with the interacting photon propagator
\begin{equation}
\label{eq:Photon}
    D_{\mu\nu}(\mathbf{Q},\omega) = \left\{[D^{(0)}_{\mu\nu}(\mathbf{Q},\omega)]^{-1} - \Pi^{\mu\nu}(\mathbf{Q},\omega)\right\}^{-1},
\end{equation}
where $\Pi^{\mu\nu}(\mathbf{Q},\omega) = \frac{1}{Vc^2} \sum_{SS'} j^{\mu\ast}_{S'}(\mathbf{Q},\omega)\tilde{L}(\omega)j^\nu_S(\mathbf{Q},\omega)$ is a (proper) dynamical polarizability which arises from the creation of virtual excitons. By computing the retarded electron-hole correlation function according to Eq.~(\ref{eq:Woodbury}) we circumvent the need to invert $L^{-1}(\mathbf{Q},\omega)$, a matrix often on the order of $10^5\times10^5$ or larger, and are left instead with the much more tractable problem of computing $D_{\mu\nu}(\mathbf{Q},\omega)$, a $4\times4$ matrix. 
Recalling that the exciton spectral function depends only on the trace of $L(\mathbf{Q},\omega)$, the problem can be further simplified to a series of vector, as opposed to matrix, operations to compute the diagonal components alone. Taken together, these dramatic reductions in computational complexity make the process of evaluating the exciton spectral function to obtain polariton dispersions trivial once the instantaneous BSE is solved and $\tilde{L}(\omega)$ constructed.

\section{Ab initio results}\label{sec3}

\subsection{Polariton dispersions and exciton localization}

\begin{figure}[ht]
    \centering
    \includegraphics[width=\columnwidth]{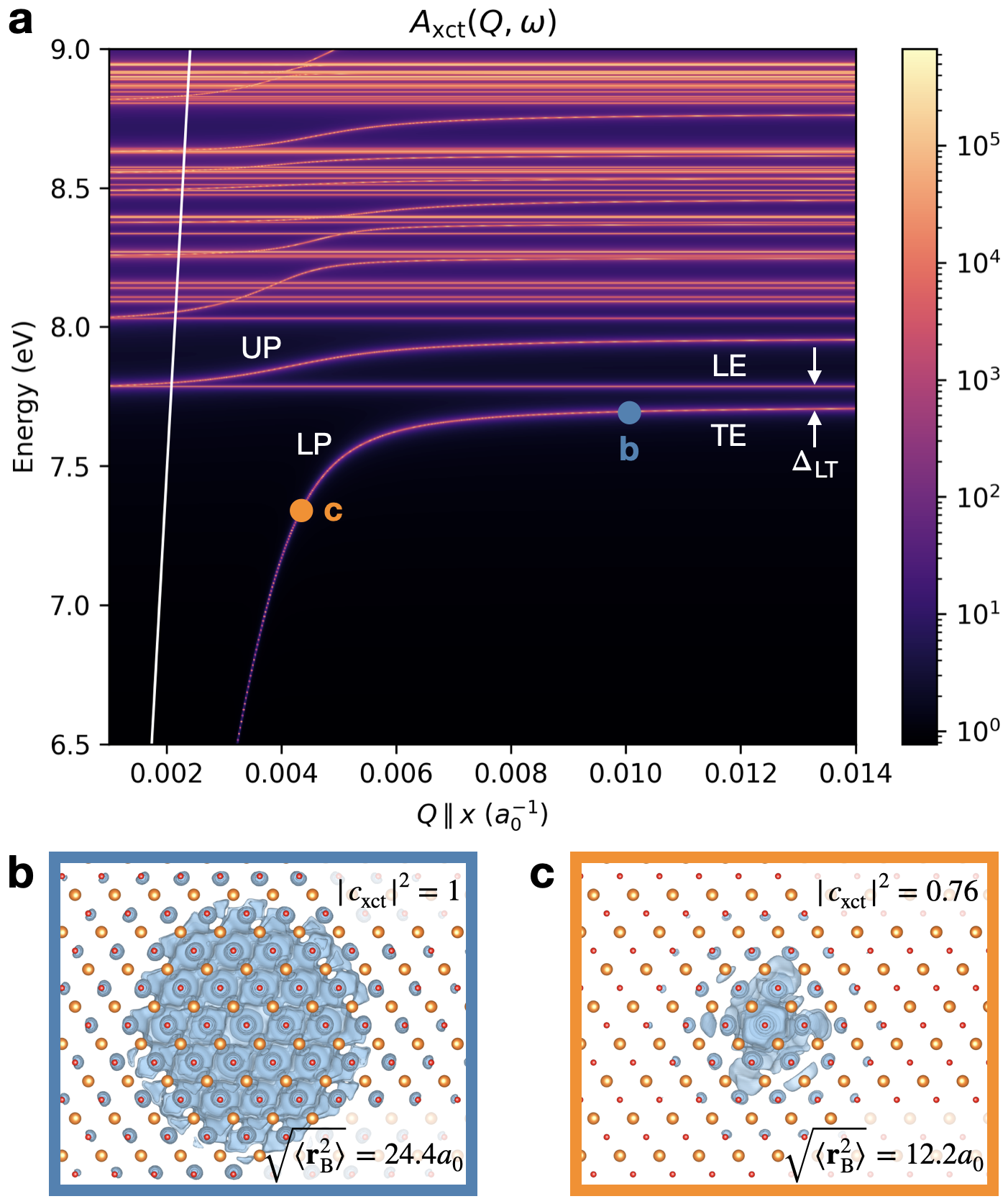}
    \caption{\label{fig:MgO} (a) Exciton spectral function of MgO, with $\mathbf{Q}\parallel\hat{\mathbf{x}}$. Only spin-singlet polaritons are shown. The LP and UP branches are composed of doubly degenerate TEs. Neither longitudinal or dipole-forbidden excitons couple to light, and appear as horizontal lines in the dispersion. (b) The excitonic portion of the polariton Dyson orbital with the hole position $\mathbf{r}_h$ fixed to be slightly shifted away from the central oxygen atom. The single-exciton fraction $|c_\mathrm{xct}|^2$ makes up the entirety of the state, indicating that it is in essence a bare exciton. (c) Same as (b) but for a photon-like LP. Attractive exchange results in a localization of the electron around the hole, shown by the reduced RMS exciton Bohr radius. The reduced single-exciton fraction is a consequence of the onset of polaritonic effects.}
\end{figure}

We first apply our formalism to the well-studied cubic semiconductor MgO by solving the instantaneous BSE for spin-singlet excitons with the BerkeleyGW package~\cite{BGW,pseudobands,SI} and using Eq.~(\ref{eq:Woodbury}) to construct the polariton spectral function (Fig.~\ref{fig:MgO}a). We observe several polaritonic branches, ranging from highly dispersive transverse polaritons to dispersionless excitons that originate from either longitudinal states or dipole-forbidden excitons (\textit{e.g.}, displaying a envelope function with $p$ or $d$-like symmetry). Importantly, instead of asymptotically approaching the bare light cone (white line) associated with the propagator $D^{(0)}$, the lower polariton branch approaches a renormalized light cone which has been screened by high-energy excitations~\cite{B02,QJL21,DT19}. This effect naturally emerges without the inclusion of any phenomenological parameters such as a high-energy background dielectric constant.

We will focus our discussion on the three lowest-energy bands in Fig.~\ref{fig:MgO}a. We first observe a lower polariton (LP) branch, which sheds its photonic character and turns into a transverse exciton (TE) as $\vb{Q}$ increases. This is followed by a longitudinal exciton (LE) that does not couple to photons, and by an upper polariton (UP). Owing to its cubic symmetry, the lower and upper polariton branches in MgO consist of doubly degenerate transverse excitons coupled to transverse photons. In the UP branch, we find that both TEs are also degenerate with the LE at $\mathbf{Q}\to0$. This is in agreement with the classical and microscopic quantum theories of exciton-polaritons for cubic crystals~\cite{DO67,BRQ86}, as well as \textit{ab initio} approaches to instantaneous electron-hole exchange~\cite{ABQ88,CAS23}.

It is instructive to understand first the effect of the \emph{static} long-range exchange interactions. If one solves the BSE as customary---without considering any long-range exchange---one obtains low-energy excitonic states displaying a three-fold symmetry and little dispersion. One may then include long-range exchange interactions in the static approximation, manifesting as a \emph{repulsive} potential on longitudinal excitations which splits the threefold degenerate excitons into two lower-energy TEs and one higher-energy LE. The associated energy difference, the LT splitting, is naturally captured by our formalism in the polaritonic dispersion far outside the light cone, wherein retardation effects in the exchange interaction are negligible (arrows in Fig.~\ref{fig:MgO}a). Close to the light cone, the coupling between TEs and photons splits these transverse states into upper and lower polariton branches. Taking this perspective, LT splitting should be interpreted as the last vestige of exciton-photon coupling in the $|\mathbf{Q}| \gg \omega/c$ limit.

A new, complementary interpretation of the polariton dispersion can be obtained through the lens of electron-hole interactions, which yields unique insights into the polariton wavefunction. The lower energy of the LP relative to the LE indicates that the retarded long-range exchange interaction effectively becomes \emph{attractive} as one approaches the light cone. To reveal the consequences of this effect, we compute the modulus square of the Dyson orbital associated with the single-exciton portion of the polariton, $|\langle 0|\psi^\dagger_{\mathbf{r}_e} \psi_{\mathbf{r}_h}|\Psi_\mathrm{Pol}\rangle|^2 = \sum_{SS'} A^{P*}_{S\mathbf{Q}} A^P_{S'\mathbf{Q}} \Phi^*_{S\mathbf{Q}}(\mathbf{r}_e,\mathbf{r}_h) \Phi_{S'\mathbf{Q}}(\mathbf{r}_e,\mathbf{r}_h)$, where $\psi^\dagger_{\mathbf{r}_e}$ ($\psi_{\mathbf{r}_h}$) is a creation (annihilation) operator associated with the electron (hole) coordinate, $\Phi_{S\mathbf{Q}}(\mathbf{r}_e,\mathbf{r}_h)$ are bare exciton wavefunctions, and $\ket{0}$ is the polariton vacuum state. Unlike the independent particle vacuum, the polariton vacuum may contain a non-zero number of bare excitons and bare photons, reflective of the quasiparticle nature of these states~\cite{QAB86}. Analyzing the excitonic component of the polariton's Dyson orbital, we observe that an attractive exchange interaction is accompanied by a significant \emph{reduction} in the exciton Bohr radius (see SI~\cite{SI}): the exciton-like polariton in Fig.~\ref{fig:MgO}b has a root-mean-square (RMS) Bohr radius of $\sqrt{\langle \mathbf{r}_\mathrm{B}^2\rangle}=24.4a_0$, whereas for the polariton in Fig.~\ref{fig:MgO}c displaying a $76\%$ single-exciton fraction, $\sqrt{\langle \mathbf{r}_\mathrm{B}^2\rangle}=12.2a_0$. This corresponds to a 50\% reduction in the relative electron-hole separation between these two states in spite of the fact that both are predominantly matter-like, possessing single-exciton fractions greater than 75\%.

\begin{figure}[ht]
    \centering
    \includegraphics[width=\columnwidth]{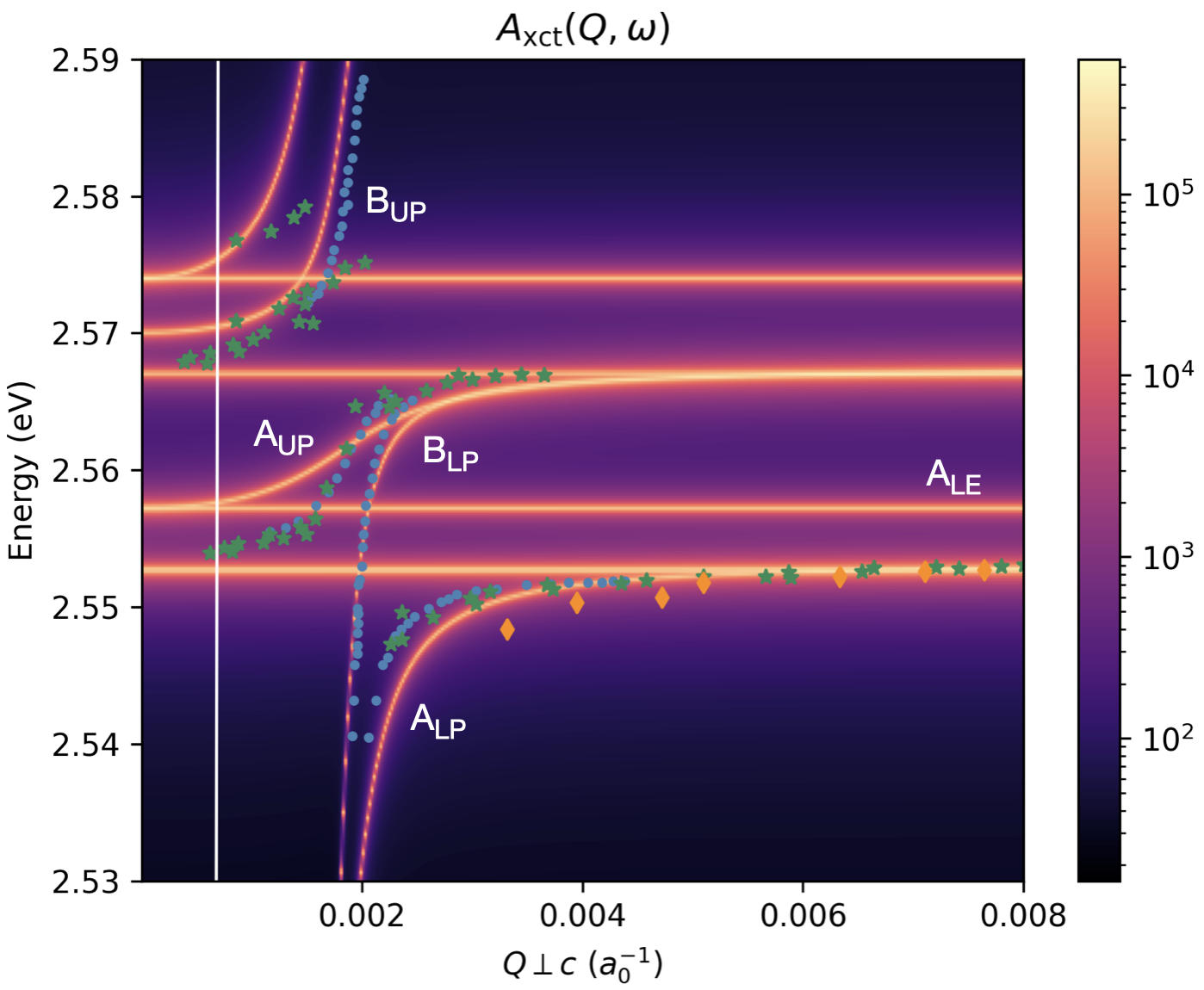}
    \caption{\label{fig:CdS} Exciton spectral function of wurtzite CdS with $\mathbf{Q}$ oriented to be orthogonal to the $\mathbf{c}$-axis of the crystal (\textit{e.g.}, along $\hat{\mathbf{x}}$). To correct for an overestimation of the electronic band gap and facilitate the comparison with experiment, a rigid shift of 102 meV has been applied to the instantaneous exciton energies. Experimentally determined points on the polariton dispersion have been overlayed as: blue circles, obtained from thin prism refraction with unpolarized light~\cite{BBBB81}; green stars, obtained from thin prism refraction with $\mathbf{E}\perp \mathbf{c}$~\cite{LTSC84}; orange diamonds, obtained from resonant Brillouin scattering~\cite{SLKH79}.}
\end{figure}

To validate our approach against available experimental data~\cite{SLKH79,LKBS82,LTSC84,KW79,BBBB81}, we next study polaritons in wurtzite CdS (Fig.~\ref{fig:CdS}). The absorption spectrum of CdS is dominated by three intrinsic exciton series A, B, C, arising from direct transitions between three distinct valence band edges to a single conduction band edge at $\Gamma$~\cite{TH59}. These three bands result from a complete lifting of the degeneracy of the three p-like valence bands at $\Gamma$ due to crystal field and spin-orbit effects. A striking feature of both the experimental dispersion curves and our result is the presence of two inequivalent LP branches: one associated with the lowest-energy optically active A exciton---polarized along $\hat{\mathbf{y}}$ due to our choice of $\mathbf{Q}\parallel\hat{\mathbf{x}}$--- and the second originating from the B exciton---polarized along the $\mathbf{c}$-axis. Their energy separation and orthogonal dipole moments manifest as a directionally dependent refractive index, a feature characteristic of birefringent wurtzite crystals. We find good agreement with all three experimental data sets, particularly regarding the screening of the polariton bands due to higher energy transitions, which our \textit{ab initio} approach naturally includes. The only notable deviation we observe is an overestimation of the computed LT splitting relative to experiments. For the lowest-energy A exciton, we find an LT splitting of $\Delta_\mathrm{LT}=4.4$ meV compared to reported experimental values of $\Delta_\mathrm{LT}=1.7\pm0.9$~meV~\cite{SLKH79,LKBS82,HT61}.
We note that our LT splitting is already overestimated within a static approximation to the exchange interactions; hence, it is likely related to the details of the underlying quasiparticle wavefunctions used in our calculations, approximated here as Kohn-Sham orbitals, and not a shortcoming of our dynamical formalism.

\begin{figure}[ht]
    \centering
    \includegraphics[width=\columnwidth]{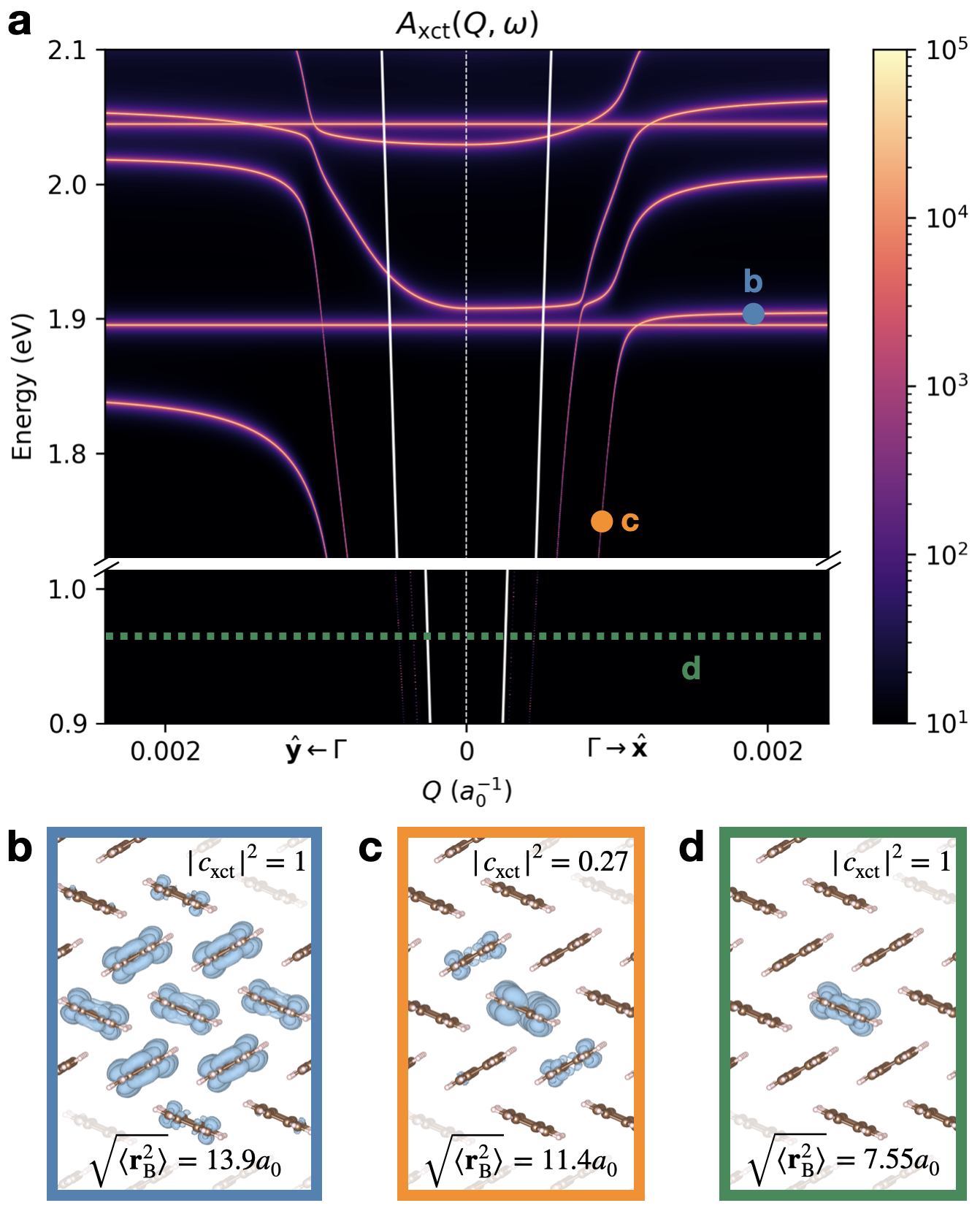}
    \caption{\label{fig:Pent} (a) Exciton spectral function of pentacene. The spectral function has been restricted to spin-singlet states, with the exception of the lowest-energy spin-triplet shown as a dashed green line. (b) Excitonic portion of the polariton Dyson orbital with the hole position $\mathbf{r}_\mathrm{h}$ averaged around the central carbon ring of a pentacene chain (see SI~\cite{SI}). The single-exciton fraction $|c_\mathrm{xct}|^2 = 1$ is equal to unity for this exciton-like point in the dispersion, indicative of the lack of polaritonic effects far from the light cone. (c) Same as (b) but for the photon-like polariton, showing reduced electron-hole separation due to an attractive exchange interaction and an associated decrease in the single-exciton fraction of this state. (d) Same as (b) but for the lowest-energy triplet exciton, which shows a characteristically Frenkel-like character.}
\end{figure}

Finally, to show the general applicability of our approach, we repeat our analysis for the more structurally complex, and non-covalently bonded molecular crystal pentacene.
Crystalline pentacene is a particularly interesting case study due to its large (short-range) exchange interactions which are known to qualitatively alter excitonic states: the lowest-energy spin-triplet excitons, which do not couple to photons  and thus do not experience exchange interactions, display a Frenkel-like character with a Bohr radius mostly confined to a single molecule~\cite{Davydov} (Fig.~\ref{fig:Pent}d). Conversely, repulsive exchange interactions lead to spin-singlet excitons which are roughly 1~eV higher in energy and display a Wannier-like character with a Bohr radius spanning several molecules (Fig.~\ref{fig:Pent}b).

Due to reduced crystal symmetry, the lowest-energy optically-active excitons in pentacene are nondegenerate with transition dipole moments which differ from each other in polarization~\cite{Davydov}. Consequently, the notion of a longitudinal or transverse exciton is restricted to polarizations along the $\mathbf{a}$-axis ($\mathbf{Q}\parallel \mathbf{a}$) which are not shown here. This lower symmetry is also responsible for breaking the degeneracy of the LP and UP branches, resulting in two distinct lower as well as upper polariton dispersions for both orientations of $\mathbf{Q}$.

We will focus for now on the LP branch of pentacene along $\mathbf{Q}\parallel\hat{\mathbf{x}}$, and plot the single-exciton component of the polariton Dyson orbital (Fig.~\ref{fig:Pent}b and c). The RMS Bohr radius of the exciton-like polariton is $\sqrt{\langle \mathbf{r}_\mathrm{B}^2\rangle}=13.9a_0$, whereas in the photon-like case we find that $\sqrt{\langle \mathbf{r}_\mathrm{B}^2 \rangle}=11.4a_0$, indicating a roughly 18\% decrease in the exciton Bohr radius and a marked reduction in spread over neighboring molecules. What is even more striking, however, is that this spin-singlet polariton, which previously spanned about seven molecules when polaritonic effects were absent (Fig.~\ref{fig:Pent}b), now spans only about three molecules (Fig.~\ref{fig:Pent}c), demonstrating a localization nearing that of the lowest-energy spin-triplet exciton (Fig.~\ref{fig:Pent}d). This highlights how dynamical interactions with the photon field can be strong enough to qualitatively change the polariton wavefunction and overwhelm the strength of the short-range exchange interactions.

\subsection{Excitonic and photonic fractions of polaritons}

\begin{figure*}
    \centering
    \includegraphics[width=\textwidth]{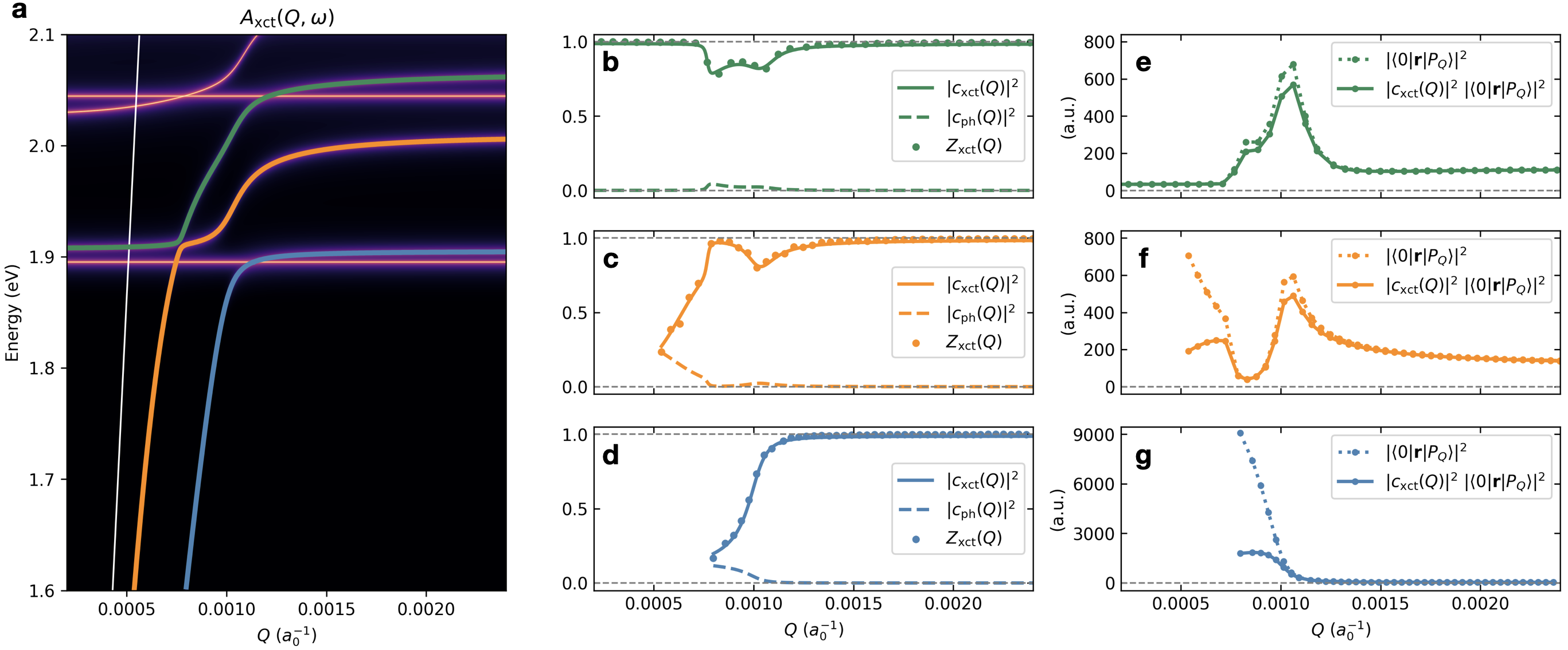}
    \caption{\label{fig:PentDipsRes} (a) Exciton spectral function of pentacene for $\mathbf{Q}$ oriented along $\hat{\mathbf{x}}$. We focus on the three highlighted bands. (b)-(d) The single-exciton (solid lines) and single-photon components (dashed lines) given by the residues of $A_\mathrm{xct}(\mathbf{Q},\omega)$ and $A_\mathrm{ph}(\mathbf{Q},\omega)$, respectively, for poles along highlighted bands. The exciton renormalization factor (dots) is also plotted as a measure of the single-exciton fraction. (e-g) The modulus square of the transition dipole moment (dotted lines) and the same quantity weighted by the single-exciton fractions (solid lines).}
\end{figure*}

In addition to providing dispersions and excitonic Dyson orbitals, our formalism also encodes information about the relative light-matter composition of polaritons. The fraction of polariton $\ket{P_\mathbf{Q}}$ composed of a single-exciton is given by the residue $\mathrm{Res}[A_\mathrm{xct}(\mathbf{Q},\omega),\Omega^P_\mathbf{Q}]$, and analogously, the residue of the photon spectral function $A_\mathrm{ph}(\mathbf{Q},\omega)=-\frac{1}{\pi}\mathrm{Im}\sum_\epsilon \epsilon^\mu D_{\mu\nu}(\mathbf{Q},\omega) \epsilon^\nu$ yields the single-photon fraction. These quantities are related to the modulus square of the Hopfield coefficients~\cite{H58}, given by $|c_\mathrm{xct}(\mathbf{Q})|^2$ and $|c_\mathrm{ph}(\mathbf{Q})|^2$ for the single-exciton and single-photon components, respectively. 

Both residues are plotted in Fig.~\ref{fig:PentDipsRes}b-d for the three lowest-energy $\mathbf{Q}\parallel\hat{\mathbf{x}}$ polariton branches of pentacene alongside the exciton renormalization factor 
\begin{equation}
\label{eq:Zfactor}
   Z^P_\mathrm{xct}(\mathbf{Q}) = \left[1 - \frac{\partial}{\partial\omega} \langle P_\mathbf{Q} | K^\mathrm{x}_\mathrm{LR} | P_\mathbf{Q}\rangle \right]^{-1}_{\omega=\Omega^P_\mathbf{Q}},
\end{equation}
as an alternate metric of the single-exciton character. For all three highlighted bands in the dispersion, we find that $|c_\mathrm{xct}(\mathbf{Q})|^2$ approaches unity at large $\mathbf{Q}$, where the polaritons are in essence bare excitons. Close to the light cone, however, the single-photon component $|c_\mathrm{ph}(\mathbf{Q})|^2$ of both LP branches rivals $|c_\mathrm{xct}(\mathbf{Q})|^2$ as the polariton begins to more closely resemble a photon. Taking the sum of these components in this photon-like regime yields a total fraction much smaller than one. This finding is a consequence of our approach's natural inclusion of the well-known multi-photon and multi-exciton character of polaritons at large coupling strengths~\cite{ConfinedElectronsPhotons}. Yet, even as the polariton dispersion approaches the light cone we find that the single-exciton fraction is relatively robust, hinting at the potential for polaritonic physics to be used to enhance desirable optoelectronic properties without sacrificing appreciable excitonic character.

One such property is the exciton oscillator strength, which plays a crucial role in light-matter coupling and exciton scattering rates. In Fig.~\ref{fig:PentDipsRes}e-g we plot the modulus square of the polariton transition dipole moment $|\langle 0 | \mathbf{r} | P_\mathbf{Q}\rangle|^2$ as a proxy for the oscillator strength (see SI~\cite{SI}). Due to a reduced exciton Bohr radius in the presence of an attractive exchange interaction, we predict an order of magnitude increase in the exciton oscillator strength in the polaritonic regime of the dispersion. We can additionally obtain a quantity which dictates the efficiency of polariton scattering mechanisms, well-known to depend on the degree of excitonic character~\cite{GOPM09, LPL21, TTST18}, by weighting $|\langle 0 | \mathbf{r} | P_\mathbf{Q}\rangle|^2$ with its single-exciton fraction. Again, we observe a remarkable enhancement in the now-weighted exciton transition dipole moment, with a maximum which is no longer in the predominantly photon-like regime for both LP branches. By probing such regions of the dispersion, one could potentially obtain oscillator strengths far larger than that of a bare exciton, while still corresponding to states with a significant, if not majority, excitonic character.

\section{Conclusions}\label{sec4}

In summary, we have presented a formalism for treating light-matter coupling and exciton-polaritons from first principles in a computationally efficient method by extending the GW plus Bethe-Salpeter equation approach to include retardation effects in the electron-hole exchange interactions. Our approach not only captures a renormalization of the exciton dispersion, which was shown to be in excellent agreement with experimental results in CdS, but also enables the real-space visualization of the excitonic component of exciton-polariton wavefunctions, revealing a dramatic reduction in the relative electron-hole separation due to the attractive nature of dynamical exchange interactions. We have also quantified the single-exciton and single-photon fractions of polaritons in different regimes of coupling strength, revealing a surprisingly robust excitonic character even for polaritons close to the light cone. Altogether, our general framework is well-suited to enable future predictive studies for utilizing polaritonic effects to engineer new electronic states, transport properties, and optoelectronic properties of materials.

\bibliography{main}

\end{document}